\documentclass[12pt]{iopart}
\usepackage{amssymb}
\usepackage{iopams,bm,verbatim,enumerate}
\usepackage[T1]{fontenc}
\usepackage{amsthm}

\def\Journal#1#2#3#4{{#4} {\it #1} {\bf #2}, #3 }

\def\td{\tau'}
\def\tdbar{\overline{\tau}'}
\def\tbar{\overline{\tau}}
\def\rd{\rho'}

\def\phidbar{\overline{\phi}'}
\def\phibar{\overline{\phi}}
\def\Psitwobar{\overline{\Psi_2}}
\newcommand{\w}[1]{\bm{#1}} 

\def\tho{\textrm{\TH}}
\def\thd{\tho '}
\def\et{\eth}
\def\etd{\eth '}
\def\ud{\textrm{d}}
\def\U{\mathcal{A}}
\def\u{U}
\def\v{V}
\newcommand{\be}{\begin{equation}}
\newcommand{\ee}{\end{equation}}

\begin{document}
%
\title{Hypersurface homogeneous Killing spinor space-times}

\author{N.~Van den Bergh}

\address{Ghent University, Department of Mathematical Analysis IW16, \\ Galglaan 2, 9000 Ghent, Belgium}
\eads{\mailto{norbert.vandenbergh@ugent.be}}

\begin{abstract}
      I present a complete list of hypersurface homogeneous space-times admitting a non-null valence two Killing spinor, including a new class admitting only exceptional Killing 
      tensors.  A connection is established with the classification of locally rotationally symmetric or boost symmetric space-times.
\end{abstract}

\pacs{04.20.Jb}

\section{Introduction}
This paper is a continuation of the results on homogeneous KS space times obtained in \cite{KShom}. KS space-times were defined \cite{McLenVdB1993} as non-conformally flat 
space-times $(\mathcal{M}, \mathbf{g})$, admitting a non-null valence two 
Killing spinor $\mathbf{X}$ or, equivalently, as the Petrov type D conformal Killing-Yano space-times. The square of their conformal Killing-Yano
two-form is a conformal Killing tensor $\mathbf{K}$ of Segre type $[(11)(11)]$.
In the conformal representant in which the Killing spinor is of modulus one, the so called \emph{unitary 
representant}, $\mathbf{K}$ is necessarily an \emph{exceptional} Killing tensor, i.e.~a non-trivial ($\mathbf{K} \not \sim \mathbf{g}$) Killing tensor 
possessing two constant eigenvalues. The classification of the corresponding space-times depends crucially on the existence of conformal representants admitting 
\emph{regular} or \emph{semi-regular} Killing tensors (these being defined as Killing tensors admitting two non-constant eigenvalues or a single non-constant eigenvalue respectively), as then the eigenvalues 
can be used to set up a preferred coordinate system. The regular case, which includes a wide range of physically interesting metrics, was dealt with successfully in \cite{Jeffryes1984}, where
also a classification of the KS space-times was presented, based on the properties of the spin coefficients in a Weyl-aligned Geroch-Held-Penrose (GHP)-tetrad for the unitary 
representant: 
they were said to be of class 1, $1_N$, 2, 3, $3_N$ or 4 according to whether
$\rho \rho' \tau \tau' \neq 0$ (class 1), $\rho \tau \tau' \neq 0 = \rho'$ (class $1_N$ ), $\tau \tau' \neq 0 = \rho=\rho'$ (class 2), $\rho \rho' \neq 0 = \tau=\tau'$
(class 3), $\rho \neq 0 = \rho'=\tau=\tau'$ (class $3_N$) or $\rho=\rho'=\tau=\tau'=0$ (class 4). Classes 1 and 3 belong to the Robinson-Trautman family of algebraically special 
space-times, all others
belong to the Kundt family. Semi-regular and exceptional Killing tensors can only occur~\cite{McLenVdB1993} in classes $1_N$ or 1. The possibility of semi-regular Killing tensors
in class $1_N$ was first noticed in \cite{McLenVdB1993}, the resulting space-times being
hypersurface homogeneous or homogeneous~\cite{KShom}. Class 1 space-times with semi-regular Killing tensors were presented in \cite{KS1_2} and all admitted
at most a one-dimensional isometry group. Introducing the functions $\phi, \phi'$ defined by
\begin{equation}\label{defPhi01} 
  \td \Phi_{01}=-3 \rho \tau \td-2\rho \phi , \, \tau \Phi_{21}=-3 \rho' \tau \td-2\rho' \phi' ,
\end{equation}
class 1 was further sub-divided in classes $KS_1$ and $KS_2$, characterised by\footnote{There was a print error in equation (40) of \cite{KS1_2}.}
\begin{eqnarray}
KS_1:  \phi+\overline{\phi} = \phi'+\phidbar = 0, \label{condKS1}\\
KS_2:  \phi+\overline{\phi}'=0 : \label{condKS2}
\end{eqnarray}
the regular class 1 metrics discussed in \cite{Jeffryes1984} made up the set $KS_1\cap KS_2$, while the metrics of \cite{KS1_2} exhausted the symmetric difference $KS_1 \Delta KS_2$. 
The existence of KS space-times belonging to $\textrm{class 1\,} \setminus (KS_1 \cup KS_2)$ (hence admitting only exceptional Killing tensors) was 
demonstrated in \cite{Beke_et_al2011}. They were found by investigating the purely Weyl-electric metrics in the class $KS_3 \supset KS_2$, characterised by
\begin{equation}
KS_3: \Im (\phi-\phi')=0. \label{KS3cond}
\end{equation} 
Note that $KS_3\cap KS_1 =  KS_2 \cap KS_1$. In \S 2 it will be shown that the extra elements
of $KS_3 \setminus KS_2$ are precisely the hypersurface homogeneous\footnote{Obviously hypersurface homogeneity is not a conformal property, but the name `hypersurface homogeneous 
KS space-times' appearing in the title may be justified as follows:  
if a Petrov type D conformal representant is hypersurface homogeneous then (because
 of the alignment of the Killing spinor with the Weyl spinor) the modulus of the Killing spinor is a 
geometric invariant and is therefore constant on the surfaces of homogeneity, such that also the unitary representant is hypersurface homogeneous. Vice versa, when the
unitary representant is hypersurface homogeneous then any conformal factor which is constant on the hypersurfaces of homogeneity will generate a conformal representant
which is hypersurface homogeneous too.} members of class 1 (including the homogeneous sub-family discussed in \cite{Beke_et_al2011}). The metrics will be explicitly constructed in \S 3.

KS space-times of classes 2, 3, $3_N$ and 4 are~\cite{Jeffryes1984} conformally related to Killing-Yano space-times. All these have been described in \cite{DietzRudiger2} and
their hypersurface homogeneous members can in principle be found by suitably restricting the Killing-Yano metrics.
However it is more convenient to construct the complete family of hypersurface homogeneous KS space-times in a coordinate invariant way
and the resulting hypersurface homogeneous metrics are presented in \S 5, 6 and 7.


Notations and conventions follow \cite{Beke_et_al2011}, which is based on the Geroch-Held-Penrose~\cite{GHP} and NP formalisms~\cite{NP}, as presented in \cite{Kramer}: 
the GHP weighted operators $\tho,\thd, \et, \etd$ generalise the NP operators 
$D,\Delta, \delta, \overline{\delta}$, corresponding to the basis one-forms $(k^a,  \ell^a, m^a,\overline{m}^a)$, while the GHP variables $\kappa',\sigma',\rho', \tau'$ 
replace the NP variables $-\nu,-\lambda, -\mu, -\pi$. All calculations were done with the aid of the Maple symbolic algebra package\footnote{A set of special purpose routines for the 
GHP and NP formalisms can be obtained from the author.} and the properties of the obtained metrics were checked with Maple's DifferentialGeometry package~\cite{Anderson_Torre}.

\section{Main equations}
I first present the main equations describing a KS space-time; for the details the reader is referred to \cite{Jeffryes1984, KS1_2}.

Choosing the spinor basis such that  \begin{equation} \label{eq1}
 X_{AB}=X o_{(A} \iota_{B)},
 \end{equation} 
the Killing spinor equation, 
\begin{equation}
\nabla _{A'(A}X_{BC)}=0 .
\end{equation}
reduces to
 \begin{equation}\label{eqKSa}
\kappa=\sigma=0
\end{equation}
and
\begin{eqnarray}
\tho X = -\rho X, \label{KSb1}\\
\et X = -\tau X. \label{KSb2}
\end{eqnarray}
These equations are accompanied by their `primed versions', namely (as $X'=X$) $\kappa'=\sigma'=0$ and $\thd X =-\rd X$, $\etd X = -\td X$.
The Weyl tensor is then of Petrov type D (or O) and $\w{k},\w{\ell},\w{m},\overline{\w{m}}$\footnote{$k^a = o^A \overline{o}^{\dot{B}}, \ell^a =\iota^A \overline{\iota}^{\dot{B}}, 
m^a = o^A \overline{\iota}^{\dot{B}}$} are its principal null directions ($\Psi_0=\Psi_1=\Psi_3=\Psi_4=0$).

In the unitary representant $|X|$ is constant and hence
\begin{equation}\label{unitary}
 \rho+\overline{\rho} = \tau+\tdbar =0.
\end{equation}

Herewith one obtains\\

\noindent
a) the integrability conditions expressing the existence of the Killing spinor:
\be \label{specKS}
\thd \rho -\tho \rd =0 , \ \et \td -\etd \tau =0, \ \tho \td -\etd \rho = 0,
\ee
b) the GHP equations:
\begin{eqnarray}
\tho \rho =0, \label{thrho}\\
\et \rho = 2 \rho \tau +\Phi_{01},\label{etrho}
\end{eqnarray}
\begin{eqnarray}
\tho \tau = 2 \rho \tau + \Phi_{01}, \label{thtau} \\
\et \tau = 0, \label{ettau} 
\end{eqnarray}
\be
\tho \rd -\et \td
= -\rho \rd -\tau \tbar-\Psi_2-\frac{1}{12} R ,\label{thrd_ettd}\\
\ee
\begin{eqnarray}
\label{defPhi00}\Phi_{00}=-\rho^2, \Phi_{02}=-\tau^2,\\
\label{def_E} E=-\frac{R}{12}-\rho \rd +\tau \td,
\end{eqnarray}
$E$ being the real part of $\Psi_2=E+i H$, and \\

\noindent
c) the Bianchi equations:
\begin{eqnarray}
\tho \Phi_{01} = -\rho (4 \Phi_{01}+5 \tau \rho), \label{bi1}\\
\thd \Phi_{01} = \rd \Phi_{01}-\rho \Phi_{12} +\tau(3 \overline{\Psi_2}+\tau \tbar-2 \Phi_{11})+\et(\Psitwobar-\frac{R}{24}+\Phi_{11}), \label{bi2} \\
\et \Phi_{01} = -\tau(4 \Phi_{01}+5 \tau \rho), \label{bi3} \\
\etd \Phi_{01}= -\tbar \Phi_{01} -\tau \Phi_{10} -\rho (3\overline{\Psi_2}+\rho \rd+2 \Phi_{11})-\tho(\Psitwobar-\frac{R}{24}-\Phi_{11}), \label{bi4} \\
\tho(\mu^2) = 3 \rd(\overline{\Psi_2}-\Psi_2) -3 \tbar \Phi_{12}-3\tau \Phi_{21}+\thd(4 \Psi_2-2 \Psitwobar+\frac{R}{24}-\Phi_{11}), \label{bi5} \\
\et(\tbar^2) = 3\tbar(\overline{\Psi_2}-\Psi_2)+3 \rd \Phi_{10}-3 \rho \Phi_{21}-\etd(\Psi_2+\Psitwobar+\frac{R}{24}+\Phi_{11}) . \label{bi6}
\end{eqnarray}

All these equations must be read as being accompanied by their `primed' and complex conjugated analogues. They can be simplified by introducing the (real) 0-weighted quantities $\u$ and $\v$, defined by
\begin{eqnarray}
 R=8(\u-\v)-16\rho \rd, \label{def_R} \\
\Phi_{11}=\u+\v-2 \rho \rd, \label{def_Phi11}
\end{eqnarray} 
($\u'=\u$, $\v'=\v$). One has then, by (\ref{def_E}),
\begin{equation}
 \Psi_2=\frac{1}{3}(\rho \rho'-2 \u +2 \v)+\tau \tau'+i H.
\end{equation}

\section{Hypersurface homogeneous KS space-times of class 1}

\subsection{General properties}
It is advantageous here to define, besides $\phi$ and $\phi'$, new 0-weighted variables $w$ (real) and $\zeta$ (complex), with $w'=w$ and $\zeta'=\overline{\zeta}$, by 
\begin{eqnarray}
\tho \rd = \thd \rho = -i w , \label{eerste}\\
\thd \phi = \rd( 2\frac{\phi \phi'}{|\tau|^2}+\zeta)+i \frac{w\phi}{\rho}.
\end{eqnarray}
Herewith one rewrites the Bianchi equations in the following form:
\begin{eqnarray}
\tho \phi = \frac{2\rho}{|\tau|^2}(|\tau|^4-|\phi|^2), \label{specbi1}\\
\et \phi = -\frac{1}{\td}\left(2 |\tau|^4+i\phi(w-H)-2\phi^2\right),\label{specbi2} \\
\etd \phi = -2\frac{|\phi^2|}{\tau}-\td(2\phi+2\overline{\phi}+\rho \rd -2 \v +\frac{i}{2 \rho} \tho H), \label{specbi3}\\
\tho \u =\rho(\phi-\overline{\phi}-3 i H), \label{specbi4} \\
\et \v = \frac{\rho\rd}{\td}(\phi-\phidbar)+i\tau(H+2 w), \label{specbi5}\\
\et H = 2 i \tau(|\tau|^2+2 \u -4\rho \rd)-2 i \frac{\rho \rd}{\td}(2 \phi+2\phidbar +\zeta) ,\label{speclaatste}
\end{eqnarray} 
together with their primed and complex conjugated analogues (taking into account $\rd = -\rho$ and $\td = -\overline{\tau}$).
Applying the commutators involving $\thd \rho$ and using (\ref{eerste}) yields two further relations, namely
\begin{eqnarray}
\tho w = 2 i \rho (\rho\rd+2\phi +2\phibar-2 \v -2 |\tau|^2), \label{thow_eq}\\
\et w = -2 i \frac{\rho \rd}{\td}(2|\tau|^2+\zeta).\label{etw_eq}
\end{eqnarray}
 Herewith all `first level' integrability conditions on $\rho,\rd,\tau,\td$ are identically satisfied. It is now easy to investigate the relation between hypersurface homogeneity, the 
 electric and magnetic parts of the Weyl tensor and the functions $\phi, \phi'$.
 
First note that in a hypersurface homogeneous space-time all 0-weighted GHP quantities, such as $\rho \rd, \tau \td, \u, \v, w$ and $H$ are functions of the 0-weighted scalar\footnote{By 
(\ref{KSb1},\ref{KSb2}) $X$ is a constant in class 4 only.} 
$X$. This suggests to define real and 0-weighted scalars $r$ and $m$ ($r,m > 0$) by
\begin{eqnarray}
 r^2 = Q \rho \rd \ \  (Q=\pm 1), \label{defr2} \\
 m^2 =|\tau|^2 .
\end{eqnarray}
Expressing that $r_{[,a} X_{,b]}=m_{[,a} X_{,b]}=0$ implies
\begin{eqnarray}
 H-w+2 i (\phibar-\phi)=0, \label{Hw_eq}\\
 Q m^2 w -2 i r^2 (\phi-\phidbar)=0,\\
 Q m^2 w - 2 i r^2 (\phi -\phibar)=0, 
\end{eqnarray}
from which one immediately infers $\phi'=\phi$. Denoting with $\mathcal{H}_i$ the set of hypersurface homogeneous space-times of class $i$ ($i=1,1_N,2,3,3_N,4$), it follows 
that $\mathcal{H}_1\subset KS_3$.\\

We now show that $KS_3\setminus KS_2 \subset \mathcal{H}_1$: evaluating the imaginary part of $\et (\ref{KS3cond})$ with (\ref{bi2},\ref{bi3}), one finds
\begin{equation}
 \left( H-w+2 i (\phibar -\phi)\right) \left( \phidbar+\phi \right)=0 ,
\end{equation}
such that, provided solutions do not belong to $KS_2$, again (\ref{Hw_eq}) follows.
Herewith (\ref{thow_eq}) implies 
\begin{equation}
 \tho H = 4 i \rho( m^2+\frac{Q}{2}r^2+\phi+\phibar-\v -\frac{2}{m^2}|\phi|^2) ,
\end{equation}
together with a similar equation for $\thd H$, with which the real part of $\et (\ref{KS3cond})$ simplifies to $(\phidbar+\phi)(\phidbar-\phibar)=0$ and hence
\begin{equation} \label{phidashphi}
 \phi ' =\phi.
\end{equation}
Acting with the operators $\tho, \thd$ on (\ref{phidashphi}) gives then
\begin{eqnarray}
 H =-2 i \frac{Q r^2+m^2}{m^2}(\phibar-\phi) , \\
 \zeta = \frac{2}{m^2}(m^4-2 |\phi|^2),
\end{eqnarray}
the $\et, \etd$ derivatives of which lead to
\begin{eqnarray}
\u=3 Q r^2+\frac{3}{2}m^2-\frac{Q r^2}{m^2}(\phi+\phibar)-2\frac{|\phi|^2}{m^2} , \\
\v =-\frac{3}{2} Q r^2-m^2+\phi+\phibar +2 \frac{Q r^2}{m^4} |\phi|^2 .
\end{eqnarray}
All 0-weighted quantities (and hence all invariants) become then algebraic functions of $m,r$ and $\phi$, with 
\begin{equation}
d r = \frac{r(\phi-\phibar)}{m^2} d \log X,\ d m  = \frac{\phibar -\phi}{m} d \log X ,\ d \phi = 2\frac{|\phi|^2-m^4}{m^2} d \log X, \label{drdmdphi}
\end{equation}
implying that the corresponding space-times are hypersurface homogeneous.

Herewith we have demonstrated that the hypersurface homogeneous members of class 1 are precisely the solutions for which $\Im (\phi-\phi')=0\neq \phi ' + \phibar$, together with the 
hypersurface homogeneous members of the regular family $KS_1 \cap KS_2$. Denoting the latter as $\mathcal{H}_{1,r}$ we have: \\ 

\noindent \emph{Property 1}
 $$ \mathcal{H}_1= (KS_3\setminus KS_2) \cup \mathcal{H}_{1,r}.$$

In \cite{KShom} we showed already that the purely Weyl-electric members of $KS_3$ are (space-time) homogeneous. It is easy to see now that a stronger result holds:\\

\noindent \emph{Property 2}\\

\emph{Purely Weyl electric KS space-times of class 1 belong to $KS_3$ and are homogeneous.}\\

Proof: substitute $H=0$ in (\ref{speclaatste}), to obtain an expression for $\zeta$,
\begin{equation}
 \zeta= 4 m^2-Q\frac{m^2}{r^2}(2 \u +m^2)-2(\phi+\phidbar).
\end{equation}
This enables one to simplify the $[\tho,\, \thd] \phi$ commutator relation to
\begin{equation}
 m^2 E (\phi+m^2)=0,
\end{equation}
implying, for non-conformally flat space-times, that $\phi$ is real. Similarly $\phi'$ is real and the $KS_3$ condition (\ref{KS3cond}) holds trivially.\\

With a bit of extra work we can also demonstrate\\

\noindent \emph{Property 3}\\

\emph{Purely Weyl-magnetic solutions of class 1 do not exist.}\\

Proof: putting $E=0$ in (\ref{def_E}) yields, 
\begin{equation}
 \v-\u=\frac{3}{2}{m}^{2}-\frac{1}{2}Q{r}^{2},
\end{equation}
which allows one to obtain, together with (\ref{specbi4},\ref{specbi5}), all derivatives of $\u$ and $\v$:
\begin{eqnarray}
\et \u =\frac{1}{2}i \tau (5 H + w),\\
\tho \v = -\frac{1}{2}\rho\left( 8(\phibar-\phi)+6 i H-i w\right).
\end{eqnarray}
Applying the $[\etd,\, \et]$ and $[\et,\, \tho]$ commutators to $\u$ leads then to expressions, the imaginary part of which reduces to the $KS_1$ condition (\ref{KSb1}) and the real part of which
results in
\begin{equation}
\tho H = 2 i \rho\left( 4 Q r^2-m^2-Q\frac{r^2}{m^2}(\zeta+2\phi +2\phidbar)-2 \u+2i \frac{\phi H}{m^2}\right).
\end{equation}
Taking the $\tho$ and $\et$ derivatives of (\ref{KSb1}) shows that also the $KS_2$ condition (\ref{KSb2}) holds and that
\begin{equation}
 \zeta = 2 m^2+i Q \frac{\phi}{r^2}(3 H-w).
\end{equation}
Herewith the $[\et,\, \thd]\phi$ commutator relation yields an expression for $w$,
\begin{equation}
 w=i \frac{m^2}{\phi}\left( 2 Q r^2-m^2-2 \u+\frac{iH}{m^2\phi}(m^4-3 \phi^2)\right),
\end{equation}
substitution of which in (\ref{thow_eq},\ref{etw_eq}) leads to an inconsistency, namely $2 m^4+\phi^2= m^4-2\phi^2=0$.

\subsection{$KS_3\setminus KS_2$} 
According to the results of the previous paragraph hypersurface homogeneous KS space-times of class 1 either admit only exceptional Killing tensors (and are then characterised by 
$\phi'=\phi$ and $\Re \phi \neq 0$), or they admit regular Killing tensors and have $\phi'=\phi$ and $\Re \phi = 0$. In both cases the explicit metrics can be found
by first translating the previously obtained invariant information into Newman-Penrose language. Aligning the tetrad as before, we will fix a boost and a spatial rotation such 
that $\rho = i Q r$ (hence $\mu=i r$) and $\tau=m$. From the Newman-Penrose equations and (\ref{defPhi00},\ref{def_E},\ref{def_R},\ref{def_Phi11}) one immediately obtains then
\begin{eqnarray}
\varepsilon=i Q \frac{r}{m}\alpha, \gamma= Q \varepsilon, \\
\beta=\alpha= \frac{1}{2 m} (m^2-2 \phibar),
\end{eqnarray}
together with
\begin{eqnarray}
R= \frac{4}{m^4} \left( Q{r}^{2}+{m}^{2} \right)  \left(5\,{m}^{
4} -4\,\phi\,\overline{\phi}-2(\phi+\phibar){m}^{2} \right) , \\
\Psi_2 = \frac{4}{3m^4} \left( \phi+2\,{m}^{2} \right)  \left(\phibar -{m}^{2} \right)  \left( Q{r}^{2}+{m}^{2} \right). \label{Psi2_expr}
\end{eqnarray}
Note that by (\ref{drdmdphi}) $r/m$ and $\Re \phi$ are constants. We therefore put
\begin{eqnarray}
 m=k r,\\
\phi  = k^2 (-Q k^2 r_0^2+i \psi) , \label{phidef}
\end{eqnarray}
with $k, r_0$ constants ($k>0$ and $r_0$ real or imaginary) and $\psi$ a real function of $r$. Integrating (\ref{drdmdphi}c) gives then
\begin{equation}\label{psi_r}
 (Q\psi - Lr)(Q\psi+L r)=(r_0^2-r^2)(r^2-k^4 r_0^2),
\end{equation}
with $L$ a constant of integration, while (\ref{drdmdphi}a) gives
\begin{equation}\label{dr_def}
 d r = 2\psi \left( ik (\w{\omega}^1-\w{\omega}^2)+\w{\omega}^3-Q \w{\omega}^4\right).
\end{equation}
As $\psi$ is real, the positivity of the right hand side of (\ref{psi_r}) implies that only the following cases can occur:
\begin{enumerate}[a)]
\item $0\neq r_0$ and $L$ are real and arbitrary, the coordinate domain in which $\psi$ (and hence $\U$) is real, being the interval $]r_-,r_+[$ with
\begin{equation}\label{eqcasea}
 2 r_\pm^2 = r_0^2(1+k^4)+L^2\pm [(r_0^2(1+k^4)+L^2)^2-4k^4r_0^4]^{1/2},
\end{equation}
\item $0\neq r_0$ is imaginary, $0\neq L$ is real and
\begin{equation}\label{eqcaseb}
|L/r_0| > 1+k^2,\ \ r\in ]r_-,r_+[,
\end{equation}
\item $0\neq r_0$ is real, $0\neq L$ is imaginary and 
\begin{equation}\label{eqcasec}
\|L/r_0| < |1-k^2|,\ \ r\in ]r_-,r_+[.
\end{equation}
It is easy to check that $r_0 < r_-$ if $k>1$ and $r_+ < r_0$ if $0<k<1$.
\end{enumerate}
The case $r_0=0$ belongs to $KS_1 \cap KS_2$: this is the regular situation, which will be dealt with in \S \ref{sec3_3} .\\

Introducing new (real) basis one-forms by
\be
\fl \w{\Omega^1}=i (\w{\omega}^1-\w{\omega}^2),\ \w{\Omega}^2=\w{\omega}^1+\w{\omega}^2,\ \w{\Omega}^3=\w{\omega}^3+Q\w{\omega}^4,\ \w{\Omega}^4=\w{\omega}^3-Q\w{\omega}^4,
\ee
it follows from (\ref{dr_def}) that $\w{\Omega}^4+ k\w{\Omega}^1$ is exact:
\begin{equation} \label{Om41}
 \w{\Omega}^4+ k\w{\Omega}^1 =\frac{1}{2 \psi} \ud r.
\end{equation}
The line-element reads now
\be
\ud s^2 = 2({\w{\Omega}^1}^2+{\w{\Omega}^2}^2-Q {\w{\Omega}^3}^2+Q{\w{\Omega}^4}^2),\label{linelementOm}
\ee
and the Cartan equations become
\begin{eqnarray}
\ud \w{\Omega}^1 = 2 Q k^2 \frac{r_0^2}{r} \w{\Omega}^3 \wedge \w{\Omega}^2, \label{Cartan1} \\ 
\ud \w{\Omega}^2 = 2\frac{k}{r}(-\psi \w{\Omega}^2-Q k r_0^2 \w{\Omega}^3) \wedge \w{\Omega}^1 -2 Q k r \w{\Omega}^3 \wedge \w{\Omega}^4), \label{Cartan2} \\
\ud \w{\Omega}^3 = -2 \frac{Q}{r}(r^2 \w{\Omega}^1+k^3r_0^2 \w{\Omega}^4) \wedge \w{\Omega}^2 -2\frac{\psi}{r} \w{\Omega}^3 \wedge \w{\Omega}^4),\label{Cartan3}\\
\ud \w{\Omega}^4 = 2 Q k^3 \frac{r_0^2}{r}\w{\Omega}^2 \wedge \w{\Omega}^3.\label{Cartan4}
\end{eqnarray}
The fact~\cite{KShom} that in the homogeneous case a closed (constant) linear combination of $\w{\Omega}^2$ and $\w{\Omega}^3$ exists, suggests in the present case to look 
for a function $\U(r)$ such that $\U \w{\Omega}^2+\w{\Omega}^3$ is closed. This leads to an over-determined set of two differential equations for $\U$,
\begin{eqnarray}
 k \psi (r \U_{,r}+\U)-Q(r^2+k^2 r_0^2 \U^2)=0,\nonumber \\
 \psi(r \U_{,r}-\U)-Q k (r^2 \U^2+k^2 r_0^2)=0,
\end{eqnarray}
which, remarkably, under (\ref{psi_r}) has a unique solution, determined by
\begin{equation}
 k^2 {\U}^{2} ({r}^{2}-{r_0}^{2}) +2 Q  k
\psi \U +{k}^{4}{r_0}^{2}-{r}^{2}=0, \nonumber
\end{equation}
or, as $L$ in (\ref{psi_r}) is only defined up to sign, by
\begin{equation} \label{U_r}
 \U = \frac {Q\psi + L r}{k ( {r_0}^{2}-{{r}^{2}) }} \equiv \frac{r^2-k^4 r_0^2}{k(Q \psi-L r)}
\end{equation}
and hence
\begin{equation} \label{Ukey}
 k^2\U^2(r^2-r_0^2)+2k L r \U +r^2-k^4 r_0^2=0.
\end{equation}

Further integration depends on whether $\U$ is real or complex:

\subsubsection{$\U$ real}

$\U$ real can only occur in the cases (\ref{eqcasea},\ref{eqcaseb}) above: $L$ is real and $r_0$ is real or imaginary.
Introducing real functions $x,y$ such that $\U \w{\Omega}^2+\w{\Omega}^3= y \ud x$, the second Cartan equation becomes
\begin{equation}\label{dxdOm4}
\fl -Q k r_0^2\frac{y\U}{r \psi} \ud r \wedge \ud x -2 \frac{y}{r (k^2 \U^2-1)}\left(k^2 \psi \U^2+Q k r_0^2(k^4-1)\U+\psi \right) \ud x \wedge \w{\Omega}^4 -\ud x \wedge \ud y = 0.
\end{equation}
When the coefficient of $\ud x \wedge \w{\Omega}^4$ is non-vanishing (which happens precisely when $L\neq0$), $\w{\Omega}^4$ can then be calculated easily. Proceeding in the same way with 
(\ref{Cartan4}) to 
find $\w{\Omega}^3$, we finally obtain
\begin{eqnarray}
  -2L \w{\Omega}^4= Q k^2 z \ud x+2Qk^2r_0^2\frac{\U^2}{r \Sigma_-}\ud r+ \frac{Q}{y} \ud y, \label{Ureal_Om4} \\
  -2L \w{\Omega}^3=\left( \frac{y(r^2-k^4 r_0^2)}{kr\U} -\frac{kr\U z^2}{4r_0^2y}+\frac{Qr\U}{ky}W \right) \ud x -\frac{r\U}{2 k r_0^2 y} \ud z ,
\end{eqnarray}
with the auxiliary functions $\Sigma_\pm$ defined by
\begin{equation}
  \Sigma_\pm = k^2\U^2(r^2\pm r_0^2)\pm r^2\mp k^4r_0^2
\end{equation}
and with $W$ a function, which by (\ref{Cartan3}) is restricted to be independent of $r,y$ and $z$: $W=W(x)$. However,
as 1) the curvature components and the spin coefficients depend only on $r$, 2) the 
null tetrad is invariantly defined and 3) (\ref{Om41}) holds, the function $W$ cannot appear in the classification algorithm~\cite{Karlhede} and\footnote{This argumentation will be referred to henceforth as 
a `\emph{Cartan-Karlhede argument}'.} one can put, 
without loss of generality, $W=0$. The corresponding expressions for $\w{\Omega}^1$ and $\w{\Omega}^2$ read then
\begin{eqnarray}
  -2L \w{\Omega}^1= -Qkz \ud x-\frac{Q\Sigma_+}{k r \Sigma_-} \ud r -\frac{Q}{k y} \ud y, \nonumber \\
  -2L \w{\Omega}^2= \left( \frac{k(r^2-r_0^2)}{r}y+\frac{krz^2}{r_0^2y}\right) \ud x+\frac{r}{2kr_0^2y} \ud z.
\end{eqnarray}

When, on the other hand, $L=0$ (which requires $r_0$ to be real), the construction of the tetrad basis vectors demands the explicit integration of the Cartan equations. The procedure is standard and after repeated use of a
Cartan-Karlhede argument to eliminate free functions, one obtains
\begin{eqnarray}
 k \w{\Omega}^1=  - C \ud y -\ud z -\frac{Q k^4 r_0^2}{2} |k^4 r_0^2-r^2|^{-3/2} |r^2-r_0^2|^{-1/2}\ud r , \nonumber\\
 k \w{\Omega}^2= -Q S |r^2-r_0^2|^{1/2}\ud y +k^2\left(2 Q z|r^2-r_0^2|^{1/2} +r|k^4r_0^2-r^2|^{-1/2}\right) \ud r, \nonumber\\
 k \w{\Omega}^3= |k^4r_0^2-r^2|^{1/2} \left( -2 Qk z\ud x + \frac{Q S}{k} \ud y\right),\nonumber \\
 k \w{\Omega}^4=  k C \ud y+k\ud z+\frac{kQ}{2}r^2|k^4r_0^2-r^2|^{-3/2} |r^2-r_0^2|^{-1/2}\ud r \label{U_realsols},
\end{eqnarray}
with $S=\textrm{sin(h)} 2 k^2r_0^2 x$ and $C=\textrm{cosh(h)} 2 k^2r_0^2 x$ if $k<1$ ($k>1$). \\

Note that, because of (\ref{Ukey}) $k=1$ is only allowed if $L\neq 0$; then also $Q=1$, as otherwise the space-time is conformally flat ($\Psi_2=0$).

\subsubsection{$\U$ not real}

As remarked above, this is the case where $r_0$ is real and $0\neq L$ is imaginary. In a way, this is easier than the real case, as the one-forms $\w{\Omega}^2$ and $\w{\Omega}^3$ are
determined by taking the real and imaginary parts of $\U (\w{\Omega}^2+\w{\Omega}^3) = P (\ud x + i \ud y)$ ($P$ complex and $x, y$ real). The second and fourth Cartan equation allow then 
to find $P$ and $\w{\Omega}^4$. With $\psi$ defined by (\ref{psi_r}), the solutions become
\begin{eqnarray}
 \fl \w{\Omega}^1 = -\frac{Q}{2k \lambda\mathcal{K}}(\mathcal{K}_{,y}\ud x - \mathcal{K}_{,x} \ud y) -\frac{1}{k}\ud z - \frac{r^2}{2 k\psi(r_0^2-r^2)}\ud r,\nonumber \\
\fl  \w{\Omega}^2 = \left(\frac{k |r_0^2-r^2|}{\lambda} \right)^{1/2} \frac{\textrm{sgn}(k-1)}{\mathcal{K}}(Q \sin 2 \lambda z \ud x-\cos 2\lambda z \ud y), \nonumber\\
\fl  \w{\Omega}^3 = \left( \lambda k |r_0^2-r^2|\right)^{-1/2} \frac{1}{\mathcal{K}} \left( (\lambda r \cos 2 \lambda z+\psi \sin 2 \lambda z) \ud x +Q (\lambda r 
\sin 2\lambda z -\psi \cos 2\lambda z ) \ud y \right), \nonumber \\
 \fl \w{\Omega}^4 = \frac{Q}{2 \lambda \mathcal{K}}(\mathcal{K}_{,y}\ud x - \mathcal{K}_{,x} \ud y) + \ud z +\frac{1}{2 } \frac{r_0^2}{\psi(r_0^2-r^2)}\ud r , \label{U_compsols}
\end{eqnarray}
where $\lambda=|L|$ and $-\log \mathcal{K}$ ($\mathcal{K}=\mathcal{K}(x,y)$) is an arbitrary solution of the Liouville equation,
\begin{equation}\label{eqLiou}
 Z_{,xx}+Z_{,yy}+\textrm{sgn}(k-1) \lambda k^4r_0^2 e^{2 Z}=0.
\end{equation}
Although the general solution of (\ref{eqLiou}) is defined up to an arbitrary analytic function, again the Cartan-Karlhede argument allows one to 
put,
\begin{equation}
 \mathcal{K} =  1-\textrm{sgn}(k-1)\lambda k^3 r_0^2 (x^2+y^2).
\end{equation}

\subsection{$\mathcal{H}_{1,r}$}\label{sec3_3}

Having found all the exceptional hypersurface homogeneous KS space-times we now investigate, in view of property (1), the regular family $\mathcal{H}_{1,r}$. The relevant equations can be obtained by
putting $r_0=0$ in equations (\ref{phidef}) to (\ref{U_r}). By (\ref{psi_r}) $L$ (and hence also $\U$) is then necessarily real and $r\in ]0, L[$, while, 
by (\ref{Cartan1},\ref{Cartan4}), $\w{\Omega}^1$ and 
$\w{\Omega}^4$ are exact. This guarantees that the first term of (\ref{Ureal_Om4}) can be absorbed in the last term ($z=z(x)$) and hence 
\begin{equation}
 -2 L \w{\Omega}^4=\frac{Q}{y} \ud y .
\end{equation}
The fourth Cartan equation gives now no extra information and $\w{\Omega}^3$ has to be calculated by explicit integration of (\ref{Cartan3}). The resulting expression,
\begin{equation}
 2 L \w{\Omega}^3 = \frac{r^2 \psi}{L y (Q L^2r -Q r^3+L \psi)}\left(y^2 \ud x+(\frac{4(Lr+\psi)^2}{r^4})^Q \ud z\right),
\end{equation}
suggests the coordinate transformation  $\frac{2\psi}{r^2}=t-\frac{1}{t}$. By (\ref{psi_r}) this implies $\frac{2L}{r}=t+\frac{1}{t}$ and, after $y\rightarrow e^{Qy}$ and a rescaling by $2L$, the canonical 
null tetrad can then be written as follows:

\noindent
a) $Q=1$:
\begin{eqnarray}
 \w{\omega}^1=\overline{\w{\omega}^2} = \frac{i}{4 k}(\frac{\ud t}{t}-\ud y )-\frac{k t}{t^2+1}(\frac{e^y}{2}\ud x-2 e^{-y} \ud z ), \nonumber\\
 \w{\omega}^3 = \frac{\ud y}{4}+\frac{1}{2(t^2+1)}(e^y\ud x+4 t^2e^{-y}\ud z), \nonumber\\
 \w{\omega}^4 = -\frac{\ud y}{4}+\frac{1}{2(t^2+1)}(e^y\ud x+4 t^2e^{-y}\ud z). \label{pos1_rsols}
\end{eqnarray}

\noindent
b) $Q=-1$:
\begin{eqnarray}
 \w{\omega}^1=\overline{\w{\omega}^2} = \frac{i}{4 k}(\frac{\ud t}{t}-\ud y )-\frac{k t}{t^2+1}(\frac{e^{-y}}{2}\ud x-\frac{e^{-y}}{8} \ud z ),\nonumber\\
 \w{\omega}^3 = -\frac{\ud y}{4}+\frac{1}{2(t^2+1)}(e^{-y}t^2\ud x+\frac{e^{y}}{4}\ud z),\nonumber\\
 \w{\omega}^4 = -\frac{\ud y}{4}-\frac{1}{2(t^2+1)}(e^{-y}t^2\ud x+\frac{e^{y}}{4}\ud z). \label{neg1_rsols}
\end{eqnarray}
\ \\

For all solutions discussed in the present paragraph, i.e.~(\ref{U_realsols}, \ref{U_compsols}, \ref{pos1_rsols}, \ref{neg1_rsols}), it is clear from (\ref{Om41},\ref{linelementOm}) 
that the hypersurfaces of 
homogeneity are space-like when $Q=-1, |k| < 1$ and time-like in all other cases (note that, by (\ref{Psi2_expr}) 
the case $|k|=1, Q=-1$ is conformally flat).

\section{Hypersurface homogeneous KS space-times of class $1_N$}

When $\mu = 0$ (\ref{etrho}') implies $\Phi_{12}=0$. There are then always conformal representants~\cite{McLenVdB1993} admitting a regular or semi-regular 
Killing tensor. The semi-regular case was treated in \cite{McLenVdB1993} and \cite{KShom}: all solutions turned out to be $T_3$-homogeneous of Bianchi class VIII or (space-time) homogeneous. 
In the regular case there are no homogeneous solutions~\cite{KShom}, but hypersurface homogeneous solutions do exist: they can be constructed explicitly, along the same lines of the 
previous paragraph, and turn out to be precisely the 
regular limits (i.e.~in which the real part $\phi$ vanishes) of the space-times considered in \cite{McLenVdB1993}. Their canonical (Weyl-aligned) null-tetrad can be simplified to the following form:
\begin{eqnarray}
 \w{\omega}^1 = \overline{\w{\omega}^1} = \sin r (2 e^{-2 k y} k^2 \ud x-\frac{1}{2} e^{2 k y} \ud z)-\frac{i}{2}(\ud y+\frac{1}{2 k} \frac{\ud r}{\sin r}),\nonumber \\
 \w{\omega}^3=2k^2e^{-2 k y} (1+\cos r ) \ud x +\frac{1}{2} e^{2ky}(1-\cos r)\ud z, \nonumber\\
 \w{\omega}^4 = \ud y .\label{reg1Nsols}
\end{eqnarray}
The $r,x$ or $z=constant$ coordinate surfaces are time-like (while the $y=constant$ surface is clearly null): the space-time is $T_3$-homogeneous, with Killing vectors given by $\partial_x, \partial_z$ and 
$2 z \partial_z - 2 x \partial_x -\frac{1}{k} \partial_y$. The Bianchi type is $VI_0$.

In the next sections we briefly consider the remaining regular cases, classes $2,3, 3_N$ and $4$.

\section{Hypersurface homogeneous KS space-times of class $2$}

When $\rho=\mu=0$ the NP tetrad can be fixed, modulo a remaining boost, by requiring $\tau$ to be real. By (\ref{KSb1},\ref{KSb2}) hypersurface homogeneity implies then that the operators $D, \Delta $ and 
$\delta+\overline{\delta}$ are identically 0 (when acting on invariantly defined quantities). From the Bianchi and NP equations it follows then that $\varepsilon$ and $\gamma$
are real and $\beta=\overline{\alpha} -i \tau / (2 H)$, with
\begin{equation}
 \delta \tau = - i H/2, \  \delta H = 2 i \tau (\tau^2+2 \u) , \ \delta \v = i \tau H . \label{class2eqs}
\end{equation}
Hence
\begin{equation}
 \v+\tau^2=\v_0 = constant \label{class2eqsb}
\end{equation}
and 
\begin{equation}\label{om1min2}
 \w{\omega}^1 -\w{\omega}^2=\frac{2 i}{H} \ud \tau,
\end{equation}
with $H=H(\tau)$ an arbitrary function.
The Cartan equations show then that $\w{\omega}^3$ and $\w{\omega}^4$ are closed, allowing to partially fix a boost such that
$\w{\omega}^3$ is exact. We have then $\varepsilon=0$ and $\alpha= \tau/2 -i H/ (4 \tau)$ and the remaining Cartan equations become
\begin{eqnarray}
 \ud (\w{\omega}^1+\w{\omega}^2) = -iH/\tau \w{\omega}^1 \wedge \w{\omega}^2+4\tau \w{\omega}^3 \wedge \w{\omega}^4 ,\\
 \ud \w{\omega}^4 = -2 \gamma \w{\omega}^3 \wedge \w{\omega}^4,
\end{eqnarray}
while the NP equations reduce to $D \gamma = 2 \v_0$ and $\delta \gamma = \overline{\delta} \gamma =0$. Like for the (space-time) homogeneous situation (occurring for $H=0$, i.e.~when $\tau$
is constant), $2 \w{\omega}^3\w{\omega}^4$ is then the
metric of a two-space of constant curvature. Distinguishing the flat ($\v_0=0$) and non-flat ($\v_0\neq 0$) cases, one obtains the following solutions:\\

\noindent
a) $\v_0\neq 0$: 
\begin{eqnarray}
 \w{\omega}^1+\w{\omega}^2 = \frac{\tau}{2 \v_0}( \ud x +\frac{\ud y -\ud z}{y}) ,\label{class2_a1}\\
 \w{\omega}^3 = \ud z,\ \w{\omega}^4=\frac{1}{8 \v_0 y^2} (\ud z -\ud y) 
\end{eqnarray}
(the $\tau, x$ or $y=constant$ surfaces are time-like and the Killing vectors are $\partial_z, \partial_x, y\partial_y+z\partial_z$ 
and $(z-y)\partial_x+(y-2z)y \partial_y-z^2\partial_z$.\\

\noindent
b) $\v_0=0$:
\begin{eqnarray}
 \w{\omega}^1+\w{\omega}^2 = 4 \tau( \ud x + z\ud y) ,\\
 \w{\omega}^3 = \ud z,\ \w{\omega}^4=\ud y \label{class2_b2}
\end{eqnarray}
(the $\tau$ or $x=constant$ surfaces are time-like and the Killing vectors are $\partial_x, \partial_y, y\partial_y-z\partial_z$ 
and $y\partial_x-\partial_z$).\\

In both cases we have a boost-isotropic space-time, with, by (\ref{om1min2}), a $G_4$ on $T_3$.

\section{Hypersurface homogeneous KS space-times of classes $3$ and $3_N$}

Solutions of class $3_N$ cannot be hypersurface-homogeneous, without being also space-time homogeneous. This can be easily seen by substituting the $3_N$ conditions 
$\rho\neq0=\rho'=\tau=\tau'$ (and hence, by the GHP equations, $\Phi_{01}=\Phi_{01}'=\Phi_{02}=\Phi_{00}'$) in the 
Bianchi equations, to derive that $\tho \u = \thd \u =0$ and $\v=0$. Imposing hypersurface homogeneity by putting $\u=\u(X)$ implies then by (\ref{KSb2}) that also $\et \u =\etd \u =0$, such that
$\u$ is constant. All 0-weighted quantities are then constants and we find ourselves again in the space-time homogeneous case, which was treated in \cite{KShom}.\\

The class 3 space-times can be obtained along the same lines as in class 2 (the two being related by a Sachs transform). 
We begin by fixing the NP tetrad, modulo a rotation, by requiring $\rho = i Q t$ and $\mu = i t$, with $t$ real and $Q=\pm 1$. By (\ref{KSb2}) and hypersurface homogeneity
we have $\delta I = (D+Q \Delta) I = 0$ for all invariantly defined quantities $I$. 

From the Bianchi and NP equations one finds $\beta=-\overline{\alpha}$, together with the analogues of equations (\ref{class2eqsb}), namely $ \Re \gamma = 
Q \Re \varepsilon = Q H / t$ and
\begin{equation}\label{case3_eqs}
 D t = H/2,\ D H = 2 t (2 Q \v - t^2), \ D \u = Q t H,
\end{equation}
implying
\begin{equation}
 \u-Q t^2 = \u_0 = constant 
\end{equation}
and 
\begin{equation}
 \w{\omega}^4- Q \w{\omega}^3 = \frac{2}{H}\ud t ,
\end{equation}
with $H=H(t)$.
The Cartan equations show that $\w{\omega}^1$ and $\w{\omega}^2$ are closed, enabling one to partially fix the rotation such that $\w{\omega}^1 = P^{-1} \ud \zeta$ with
$P$ real and $\zeta$ complex. From the NP equations it follows then that $P=P(\zeta, \overline{\zeta})$ and 
\begin{equation}
 \delta \alpha-\overline{\delta}\beta + 4 \alpha \beta -2 \u_0 =0,
\end{equation}
implying that $2 \w{\omega}^1 \w{\omega}^2$ is the metric of a two-space of constant curvature and suggesting to introduce coordinates $x$ and $y$ by 
\begin{equation}
 \w{\omega}^1=\overline{\w{\omega}^2} = \frac{\ud x + i x \ud y}{1+k x^2 /4} \ (k \equiv 8 \u_0) . \label{class3_1}
\end{equation}
With $H$ an arbitrary function of $t$, the one remaining Cartan equation,
\begin{equation}
 \ud \w{\omega}^3 = 2i Q r \w{\omega}^2 \wedge \w{\omega}^1 +\frac{H}{2t} \w{\omega}^4 \wedge \w{\omega}^3, 
\end{equation}
can be integrated to yield
\begin{equation}
 \w{\omega}^3=\frac{4 Q t x y}{(1+\frac{k}{4}x^2)^2}\ud x+ t \ud r . \label{class3_3}
\end{equation}
Note that the $t=constant$ surface is time-like when $Q=1$ and space-like when $Q=-1$. The $x,y$ or $r=constant$ surfaces are always time-like. The Killing vectors are $\partial_r$, 
$\frac{k}{32}\partial_y +\frac{Q}{kx^2+4}\partial_r$ and 
\begin{eqnarray*}
 \frac{Qx}{kx^2+4}(y\cos y -\sin y) \partial_r-\frac{kx^2+4}{64} \cos y \partial_x-\frac{kx^2-4}{64} \sin y \partial_y ,\\
 \frac{Qx}{kx^2+4}(y\sin y +\cos y) \partial_r-\frac{kx^2+4}{64} \sin y \partial_x+\frac{kx^2-4}{64} \cos y \partial_y .
\end{eqnarray*}
Clearly we have a locally rotationally symmetric space-time, admitting a $G_4$ on $T_3$ or $S_3$.

\section{Hypersurface homogeneous KS space-times of class $4$}

When $\rho=\rho'=\tau=\tau'=0$ the only non-vanishing curvature components are $R,\Psi_2$ and $\Phi_{11}$, while $X$ is constant. The Bianchi identities immediately give $\tho \u = \thd \u = 0$ and $\et \v =
\etd \v = 0$, such that hypersurface-homogeneity implies that either $\u$ or $\v$ should be constants (or both, in which case we have space-time homogeneity). From the Cartan equations it follows that the four basis one-forms are closed,
enabling us to partially fix a boost and rotation such that
\begin{equation}
\w{\omega}^1 = P^{-1} \ud \zeta,\ 
\w{\omega}^3 = Q^{-1}\ud s,\ 
\w{\omega}^4 = Q^{-1}\ud t, \label{classIV}
\end{equation}
with $P$ and $Q$ real. The Cartan equations immediately imply then $P=P(\zeta,\overline{\zeta})$ and $Q=Q(s,t)$, with
\begin{equation}
 \alpha=-\overline{\beta}= \frac{1}{2}P_{,\overline{\zeta}}, \ 
 \varepsilon = -\frac{1}{2}  Q_{,t},\ \gamma= \frac{1}{2} Q_{,s} .
\end{equation}
The surviving Newman-Penrose equations are
\begin{eqnarray}
D\gamma-\Delta \varepsilon +4 \varepsilon \gamma = 2 \v ,\\
 \delta \alpha+\overline{\delta} \overline{\alpha}-4\alpha \overline{\alpha} = 2\u,
\end{eqnarray}
showing that the space-time is a product of an arbitrary two-space $\mathcal{M}_1$ and a two-space $\mathcal{M}_0$ of constant curvature (the latter's metric being $2\w{\omega}^1\w{\omega}^2 = $ if $\u$ is constant and 
$2\w{\omega}^3\w{\omega}^4$ if $\v$ is constant). Hypersurface-homogeneity implies
that also $\mathcal{M}_1$ should admit a Killing vector and hence 
the space-time is locally rotationally or boost symmetric, with metric given by 
\begin{equation}
 \ud s^2 = e[-\ud t^2+A^2(t)\ud x^2]+\frac{\ud y^2+\ud z^2}{(1+\frac{k}{4}(y^2+z^2))^2},\ (e=\pm 1) \label{Steph13_1a}
\end{equation}
or
\begin{equation}
 \ud s^2 = \ud x^2+A^2(x)\ud y^2 +\frac{\ud u \ud v}{(1+\frac{k}{4}uv)^2}.\label{Steph13_1b}
\end{equation}

\section{Discussion}

A complete classification is obtained of the (non-conformally flat) hypersurface homogeneous space-times admitting a non-null Killing spinor of valence two, i.e.~of the hypersurface-homogeneous 
Petrov type D conformal Killing-Yano space-times. In this classification two new families appear of space-times for which the associated Killing tensor is \emph{exceptional}, namely 
those determined by the metrics (\ref{U_realsols}) and (\ref{U_compsols}). Both belong to class 1 and admit a $G_3$ on $S_3$ for $Q=-1, |k| < 1$; 
for all other values of $Q$ and $k$ they admit a $G_3$ on $T_3$. 
The regular members of class I (forming a subclass of the Carter\cite{Carter68a} metrics discussed in\cite{Jeffryes1984}), are given by (\ref{pos1_rsols}, \ref{neg1_rsols}). 
The causal character of the hypersurfaces of homogeneity is identical to that of the exceptional metrics. In all cases the isometry group has Bianchi type $VI_0$. Semi-regular Killing tensors
only appear in class $1_N$ 
and were discussed in \cite{McLenVdB1993}, with all solutions being $T_3$ homogeneous of Bianchi type $VIII$. The regular members of class $1_N$ are given by the metrics (\ref{reg1Nsols}): 
all are $T_3$ homogeneous of Bianchi type $VI_0$. The hypersurface homogeneous members of class $2, 3, 3_N$ and $4$ all admit isometry-groups of dimension $>3$. The $3_N$ solutions in 
particular are
homogeneous and were discussed in \cite{KShom}. The metrics of class $3$, given by (\ref{class3_1}-\ref{class3_3}), turn out to be locally rotationally symmetric (LRS), admitting a $G_4$ 
on $T_3$ (when $Q=1$) or on $S_3$ (when $Q=-1$), while the 
class $2$ metrics (\ref{class2_a1}-\ref{class2_b2}) are their boost symmetric (BS) analogues, admitting a $G_4$ on $T_3$. Finally the class 4 metrics are either LRS or BS and admit a $G_4$ on $S_3$ or $T_3$.

All this pertains to the metrics of the \emph{unitary} representants, in which the 
Killing spinor has modulus 1 ($|K|=1$), but, the existence of a Killing spinor being a conformally invariant property and hypersurface-homogeneity being preserved under suitable 
conformal transformations, stronger conclusions hold. First note that it can be easily verified that the LRS metrics (13.1) and (13.2) of \cite{Kramer}, namely 
\begin{equation}
 \ud s^2 = e[-\ud t^2+A^2(t)\ud x^2]+B^2(t)[\ud y^2 +\Sigma(y,k)^2\ud z^2], \label{SSteph13_1a}
\end{equation}
and
\begin{equation}
 \ud s^2 = e[-\ud t^2+A^2(t)(\ud x+\sigma(y,k) \ud z)^2]+B^2(t)[\ud y^2 +\Sigma(y,k)^2\ud z^2], \label{SSteph13_1b}
\end{equation}
($e=\pm 1$; $\Sigma(y,k)=\sin y, \sinh y$ or $y$ and $\sigma(y,k)=\cos y, \cosh y$ or $y^2/2$ according to whether $k=1, -1$ or $0$) both admit a Killing spinor, with (\ref{SSteph13_1a}) belonging to 
class 4 and (\ref{SSteph13_1b}) to class 3. It follows that the LRS metrics (\ref{SSteph13_1a}) (and their BS counterparts, namely (13.14) in \cite{Kramer}) completely exhaust the 
hypersurface homogeneous
class 4 space-times, while the LRS metrics (\ref{SSteph13_1b}) and their BS counterparts exhaust the hypersurface
homogeneous classes 3 and 2 respectively. On the other hand, the third family of 
LRS space-times (of Bianchi type $V$ or $VII_h$), given by the metrics (13.3) in \cite{Kramer},
\begin{equation}
 \ud s^2 = e[-\ud t^2+A^2(t)\ud x^2]+B^2(t)e^{2x}[\ud y^2 +\ud z^2],\ (e=\pm 1) \label{SSteph13_1c}
\end{equation}
does \emph{not}\footnote{This corrects an earlier remark in \cite{KShom} that all LRS perfect fluids should admit a Killing spinor!} admit a 
Killing spinor: using a Weyl-adapted null tetrad the spin coefficients $\rho$ and $\mu$ will be real in all conformal representants, 
contradicting the condition (\ref{unitary}) which should hold in the unitary representant.

What remains to be done is to investigate the possible physical interpretation of any of the conformal representants of these solutions. While this has been partially 
successful~\cite{Beke_et_al2011,KShom, KSprods} in the homogeneous case and for the regular families~\cite{Jeffryes1984, Papacostas1983b}, close to nothing is known so far about the exceptional metrics.

\section*{References}


\providecommand{\newblock}{}

\end{document}